\documentclass[fleqn,twoside]{article}
\usepackage{espcrc2}
\usepackage{graphicx}

\title{Progress in FDC Project}
\author{Jian-Xiong Wang\address[IHEP]{Institute of High Energy Physics,
        Academia Sinica, \\
        P.O. Box 918, 100039 Beijing, China}
       }

\begin{document}
\begin{abstract}
\vspace{1pc} The FDC is a general-purpose program package for
Feynman Diagram Calculation. We outline previous successes in
calculations and focus on its recent progress about automatic 
deduction the Feynman rules for first principle model,
especially for the supersymmetric model, proper evaluation of the rates of
multi-final-particle processes and the event generators in the SM
and MSSM. A few special applications are presented. The
FDC-homepage shows an automatic translation of the FDC results
into the HTTP version.
\end{abstract}

\maketitle

\section{Introduction}
With the progress in high energy physics, it is obvious that an
automate calculation of the physical processes by the perturbative
quantum field theory becomes very useful and important.  There are
many projects for this purpose, such as the Grace\cite{grace1993},
CompHEP\cite{comphep1995}, FeyArt\&FeyCal\cite{feyart}, Diana
\cite{diana} and several others. FDC(Feynman Diagram Calculation)
is one of these projects. It was  1992 when we started to write a
program for manipulations of one-loop amplitudes
\cite{jxwang1993}. Now a more developed program can be used to
automatically deal with  any process at tree level, and since then
many applications have been made.

The paper is organized as follows. In Section 2 we outline the
basic parts of the FDC. Automatic deduction of the Feynman rules 
for first principle model is presented and especially the procedure for
the super-symmetry model is introduced in Section 3. Section 4 is
devoted to the rate evaluation of multi-final-particle processes
and building the corresponding event generators. In Section 5, a
few sub-projects for FDC applications are displayed. Finally a
summary is given in Section 6.

\section{Basic Parts of FDC}
The main parts of the FDC are shown in Fig.1. Its source programs
are mainly written in the REDUCE and Rlisp, and there is also a
Fortran library which supplies necessary basic functions. The
Feynman Diagram plotter is written in C++.

The first two parts {\it gmodel and diag} \cite{jxwang1994} were
developed in 1993, where {\it gmodel} is used to  deduce Feynman
rules for the standard model and {\it diag} is a Feynman diagram
generator which generates Feynman diagrams to any perturbative
orders. {\it psdraw} was constructed to draw the Feynman diagrams
from the output of {\it diag} into a PS file (up to two loops)
automatically in 1994.  {\it amp} was finished in 1995 to treat
the amplitude square manipulation and related Fortran source
generation for tree processes,  much improvement had been made and
other two optimized methods were introduced later. To properly
handle the cancellation among large numbers which is required by
the gauge invariance, in the calculations at tree level, a method
was proposed \cite{jxwang1996} and  realized in the FDC in 1996.
It is a very powerful method to solve the problem in which a
t-channel photon peak appears simultaneous with the gauge
invariance breaking due to the finite widths of the involved
unstable particles. In 1997, {\it kine} was developed to treat the
phase space integration for multi-final-particle processes
\cite{jxwang1997} automatically and improvements were made later.
{\it gmprocess} was added to generate multi-final-particle
processes in 1998, it was used to generate all the processes on
$e^+e^-$, $pp$ colliders  up to four final
particles\cite{jxwang1998}. To automatic deduce the Feynman rules
for the minimal super-symmetry model and the R-parity violation
model, a modified {\it gmodel} was developed in 1999$\sim$2000
\cite{jxwang2001}. {\bf\it gen\_html} was developed in 2002 to do
the translation and presentation of the FDC package results into
the {\it HTTP version}. Therefore most of the results generated by
the FDC can be easily put on the FDC-homepage:
\begin{verbatim}
www.ihep.ac.cn/lunwen/wjx/public_html
\end{verbatim}
Many other options, functions have been implemented in the FDC
since 1993.
\begin{figure}[htb]
\includegraphics[scale=0.4]{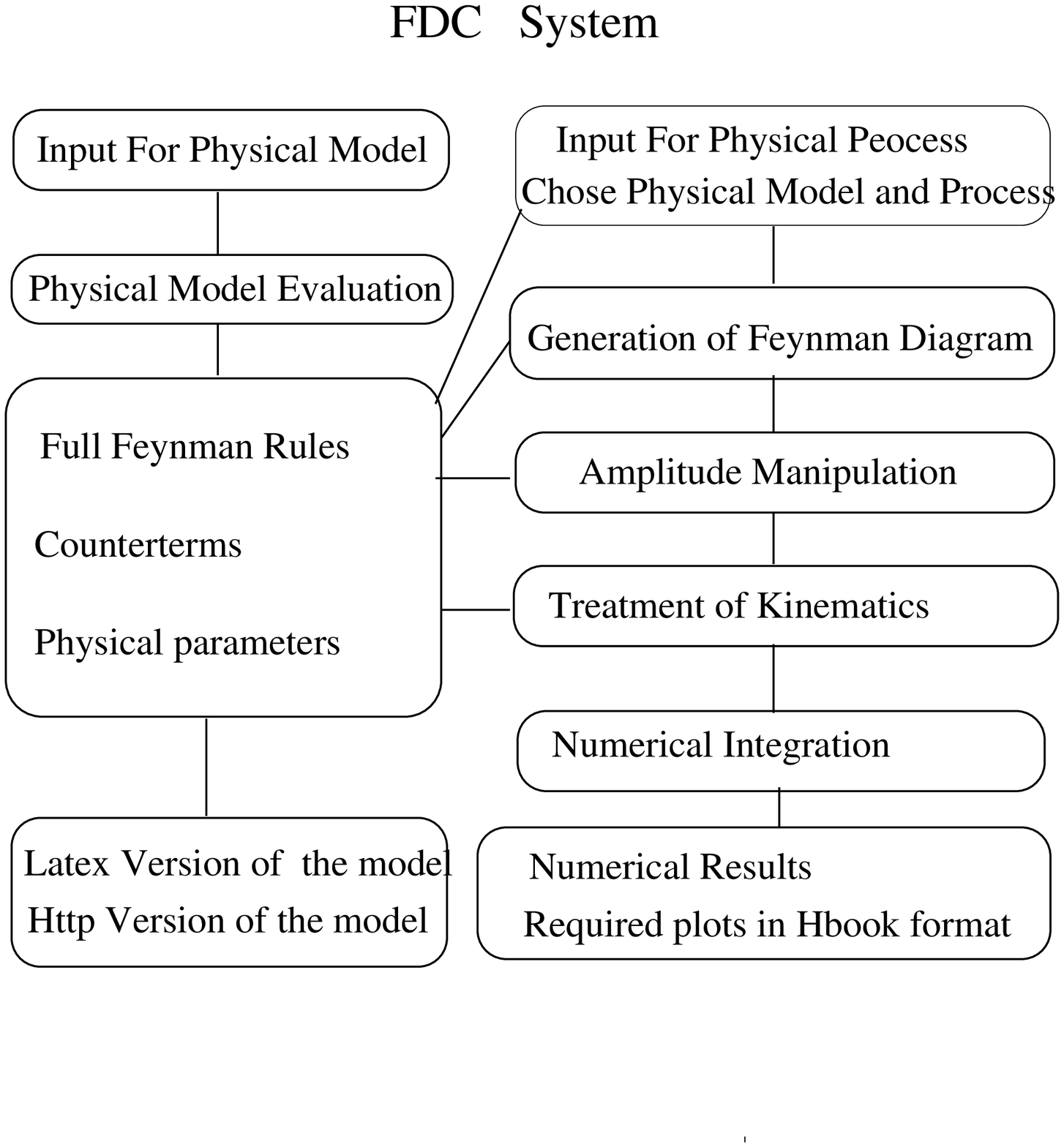}
\caption{FDC System Flow}
\label{fig:1}
\end{figure}
\section{The Supersymmetric Model}
To build up a physical model in the package is the first step. It
is done in most packages by inputting  all the Feynman rules,
counterterms and constants directly. This is not a hard task for
the Standard Model, but a very lengthy  for more complicate
models, such as the super-symmetry model.

For the first principle model, there is a standard way to
construct the Lagrangian, to quantize it and give all the Feynman
rules and counterterms. In the FDC, we have developed {\it gmodel}
to perform this standard task, i.e, to construct the Lagrangian 
and deduce the Feynman rules for the SM and the super-symmetry model.
Indeed, it is pretty easy to build in any physical model to our
established system, even though developing the program was a very
hard job.
\begin{figure}[htb]
\includegraphics[scale=0.4]{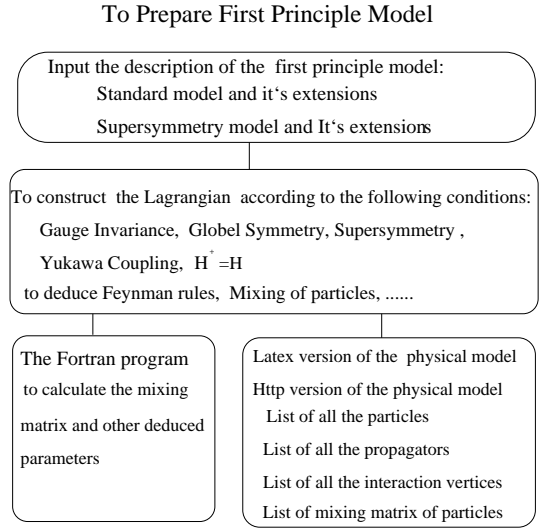}
\caption{System Flow for {\it gmodel}}
\label{fig:2}
\end{figure}
The standard procedure of including a new model in the system is
shown in Fig.2. The contents of the input documents are  very
simple and easy to understand for users, {\it gmodel} can construct
the Lagrangian and then deduce all the mixing matrices, all the 
Feynman rules from the description of the model in the input file, 
and then prepare the Latex version of the result and an internal
file for later usage in the FDC. A parameterization scheme, i.e, a
set of independent input parameters, can be chosen under interface
with users. The Fortran source is prepared to calculate the
deduced parameters which are needed in following calculations. It
has been used to generate the Feynman rules for the MSSM and the
R-parity-violated SUSY model.

It is very easy to change many things in the input file of the
model, such as the gauge fix terms, notations,  the contents of
particles, more leptons, more Higgs bosons, soft-breaking terms,
forms of the super potential, one can also choose a model which
breaks global symmetries such as the lepton number conservation,
baryon number conservation, etc.

\section{Evaluation of Multi-final particle Processes and Their Event Generators}
\begin{figure}[htb]
\includegraphics[scale=0.4]{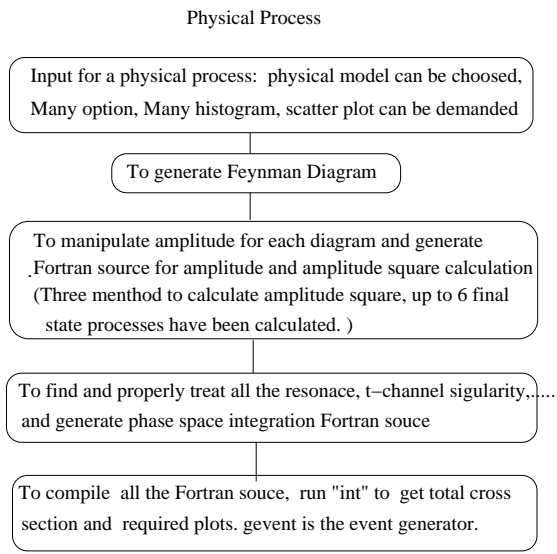}
\caption{System Flow for gprocess}
\label{fig:3}
\end{figure}
The integrated script {\it gprocess} is shown in Fig. 3. It is
used to calculate a given tree process, i.e. to prepare all the
Fortran sources for the cross section calculation and event
generator. The generator supplies an interface in the LesHouches
accord format to transfer the generated event to any parton shower
MC for the later hadronization processes.  The interface Fortran
program between the event generator and pythia\cite{pythia} is
implemented. Based on the {\it gprocess}, {\it gmprocess} is
developed to automatically generate all the processes for a
certain topic, such as
   $e^+ e^- \rightarrow 2, 3, 4~final~particles$,
   $e^+ p \rightarrow 2, 3, 4~final~particles$ and
   $p p \rightarrow 2, 3, 4~final~particles$ in the SM.

As an example, we calculate the all the decay rates of the
super-symmetry partners in this way, and compare the results with
Isajet\cite{isajet} at different Snowmass benchmark
points\cite{smbpoint2000} of the parameter space.  The event
generators are generated for all the processes with $2,3,4$
final-state particles at different collider energies, such as the
LHC, Tevatron and LC, in the SM and MSSM. A HTTP version is
written in the FDC-homepage.

\section{A Few Sub-Projects in The Application of FDC}
Recently we have extended the application of the FDC to a few very
special aspects. In these cases, many new programs as well as the
modifications of the old ones are developed. The following three
subsections are devoted to them.
\subsection{FDC-PWA}
Based on the FDC package, the sub-project FDC-PWA was completed in
1999. It was developed to generate a complete set of the Fortran
sources to do the partial wave analysis on experimental data of
decay processes at low energy regions ($J/\psi$ mass, etc.).

We have developed a new {\it gmodel} to construct effective
Lagrangians and deduce the corresponding Feynman rules based on
the basic requirements such as Isospin invariance, C-parity
invariance, P-parity invariance, G-parity invariance, Lorentz
invariance, strange number conservation, charm number
conservation, etc. The input is a list of mesons and baryons with
values of their Spins (0, 1/2, 1, 3/2, 2, 5/2, 3 , 7/2, 4, 9/2),
Isospin, P-parity, C-parity, G-parity, charm number, strange
number, masses and widths. The expression of the effective
interaction vertices and the propagators for the high spin states
are quite lengthy, and the related amplitudes and amplitude
squares are  complicated. There are many free parameters in the
effective model and these parameters will be fixed when the
generated program is used to do Likelihood fitting of experimental
data.

To work with high spin states and generate the interface to
Likelihood fitting program, there are some changes in the Feynman
diagram generator {\it diag}, the amplitude manipulation program
{\it amp} and the kinematics treatment program {\it kine}. The
FDC-PWA has been used in the partial wave analysis for the BES
experiments since 2000. The details about the FDC-PWA will be
introduced in a different paper\cite{jxwang20041}

\subsection{FDC-EMT}
A new part in FDC, FDC-EMT was developed to treat the effective
meson model \cite{liba1995} during 2001 to 2002. Part of the
results can be found in the FDC-homepage.

\subsection{FDC-NRQCD}
We have implemented the non-relativistic QCD (NRQCD) formalism for
calculating the decay  and production rates of the S-wave, P-wave,
Color octet, Color Singlet heavy quark meson states such as
$J/\psi$, $\psi^\prime$, $B_c$ in the FDC. The results about
$J/\psi$ and $B_c$ production at various colliders can be found in
the 2003, 2004 result part of the FDC-homepage. This part of the
FDC will be discussed in another paper\cite{jxwang20042}

\section{Summary}

The FDC package realizes a completely automatic deduction from
physical models to final numerical results for tree processes. We
have developed {\it gmodel} to deduce Feynman rules 
for a fwe physical models, Standard Model and their
extension, Super-symmetry model, and a few Phenomenological models
as well, i.e, NRQCD-related heavy quark meson model and Effective
Lagrangian model for Partial wave analysis. Three methods are
implemented to manipulate amplitudes and generate the the total
squared amplitudes. It can be chosen by using different options.
We have built in an automatic treatment program of phase space
integration for multi-final-particle processes. Many improvements
have been made since 1997. {\it gmprocess} has been developed for
generating  multi-final-particle processes and serves as the event
generator for such processes.  It was used to generate the event
generators for the LHC, Tevatron and LC. The {\it gen\_html} was
developed to automatically generate HTTP pages for physical models
and multi-final-particle processes. Most of the results are
presented in the FDC-homepage.

\vspace{0.5cm}
\centerline{\bf Acknowledgments}
I would like to thank Prof. Y. Shimizu, C. H. Chang, T. Ishikawa, J. Fujimoto and T. Kaneko for discussions on many related issues,
thanks to Dr. Y. Kurihara for discussion on kinematic treatment for multi-final
particles, thanks to Prof. M. Kuroda and K. Hagiwara for discussion on super
symmetry model, thanks to Prof. B. S. Zou and Y. C. Zhu for discussion on partial
wave analysis method and effective Lagrangian model, thanks to Prof. B. A. Li for
discussion on his effective meson model and thanks to Prof. C. F. Qiao for discussion
on NRQDC formalism and related issues, thanks to Prof. X. Q. Li for many discussion.
This work was supported in part by the National Natural Science foundation of China
under Grant Nos.90103013 and by the Chinese Academy of Sciences under Project
No. KJCX2-SW-N02.

\end{document}